# Alterations of the mitochondrial proteome caused by the absence of mitochondrial DNA: a proteomic view

Mireille Chevallet[1],*, Pierre Lescuyer[1,*], Hélène Diemer[2], Alain van Dorsselaer[2], Emmanuelle Leize-Wagner[2], and Thierry Rabilloud[1]

[1] CEA, Laboratoire d'Immunochimie, DRDC/ICH, INSERM U 548, CEA-Grenoble, 17 rue des martyrs, F-38054 GRENOBLE CEDEX 9, France
[2] Laboratoire de Spectrométrie de Masse Bio-Organique, UMR CNRS 7512, ECPM, 25 rue 3 Becquerel, 67008 STRASBOURG Cedex, France

* : these authors contriubuted equally to this work

Correspondence :
Thierry Rabilloud, DRDC/ICH, INSERM U 548
CEA-Grenoble, 17 rue des martyrs,
F-38054 GRENOBLE CEDEX 9
Tel (33)-4-38-78-32-12
Fax (33)-4-38-78-98-03
e-mail: Thierry.Rabilloud@cea.fr

Abbreviations:
MERRF: myoclonal epilepsy with ragged red fibers; MtDNA: mitochondrial DNA; Rho0: cells devoid of mitochondrial DNA

Abstract
The proper functioning of mitochondria requires that both the mitochondrial and the nuclear genome are functional. To investigate the importance of the mitochondrial genome, which encodes only 13 subunits of the respiratory complexes, the mitochondrial rRNAs and a few tRNAs, we performed a comparative study on the 143B cell line and on its Rho-0 counterpart, i.e. devoid of mitochondrial DNA. Quantitative differences were found, of course in the respiratory complexes subunits, but also in the mitochondrial translation apparatus, mainly mitochondrial ribosomal proteins, and in the ion and protein import system, i.e. including membrane proteins. Various mitochondrial metabolic processes were also altered, especially electron transfer proteins and some dehydrogenases, but quite often on a few proteins for each pathway. This study also showed variations in some hypothetical or poorly characterized proteins, suggesting a mitochondrial localization for these proteins. Examples include a stomatin-like protein and a protein sharing homologies with bacterial proteins implicated in tyrosine catabolism. Proteins involved in apoptosis control are also found modulated in Rho-0 mitochondria.

## 1. Introduction

Numerous pathways critical for cell life take place in mitochondria. Most of the ATP is synthetized in mitochondria, which are also a major cellular store for calcium. Important catabolic pathways such as fatty acid oxidation or citric acid cycle, and anabolic pathways such as heme synthesis, part of the pyrimidine nucleotide synthesis and coenzyme synthesis (e.g. ubiquinone) are also localized in this organelle. Moreover, mitochondria seem to play an important role in the control of apoptosis [1].

Mitochondria are among the few organelles (with plasts) to have an independent genome. The size of this genome is very variable among the various phylla. While plants have important mitochondrial genomes, the mammalian mitochondrial genomes are the smallest ones. The somewhat vestigial human mitochondrial genome encodes 13 subunits of the respiratory complexes, the mitochondrial rRNAs and some tRNAs. Despite of the importance of mitochondria in the cell life, and the fact that mutations in the mitochondrial genome often lead to severe diseases [2], cells devoid of any mitochondrial DNA, also named Rho0 cells, have been produced from transformed cells [3]. These cells grow more slowly than the original cells, and are auxotroph for pyruvate and uridine. The need for pyruvate can be explained by the fact that the cells rely entirely on glycolysis for their ATP production, and need pyruvate as an hydrogen receptor to regenerate NAD from NADH. The need for uridine can be assigned to the fact that some steps in pyrimidine biosynthesis take place in mitochondria (e.g. conversion of dihydroorotate into orotate).

Interestingly, Rho0 mitochondria are not empty organelles. While it has been shown that the absence of the mitochondrially-encoded subunits precludes the assembly and function of mitochondrial respiratory complexes [4], there is still a proton gradient in Rho0 mitochondria, which is essential for protein import into mitochondria and for prevention against apoptosis [1]. This proton gradient is generated by hydrolysing ATP imported from the cytoplasm by the ATP-ADP shuttle. ATP is hydrolyzed by the ATP synthase working the reverse way of normal cells to generate a transmembrane proton gradient [5].

As another example, Rho-0 cells do not require non-essential amino acids for their growth, showing thereby that the classical generation of these aminoacids from transamination of keto-acids takes place normally. This implies in turn that a complete citric acid cycle can take place in Rho0 mitochondria.

These physiological data told us that some processes where altered in Rho-0 cells [6], while some others were not. In order to get a better understanding of the changes induced in mitochondria upon deletion of the mitochondrial genome, we chose to perform a comparative proteomic analysis of normal and Rho0 mitochondria coming from an osteosarcoma cell line and its Rho-0 counterpart obtained by cultivation with ethidium bromide [7]. As previous work had shown that alterations in the mitochondrial proteome induced by alterations in the mitochondrial genome were often discrete and quantitative changes [8], [9], we decided to use a two dimensional electrophoresis-based approach to characterize the modifications in the mitochondrial proteome induced by the absence of mitochondrial DNA.

## 2. Materials and methods

### Cell culture and reagents

Both 143B osteosarcoma cells and their Rho0 counterpart were a gift from G. Attardi (Caltech). Both cell lines were cultured in high glucose DMEM, supplemented with 1mM sodium pyruvate and 0.4mM uridine. Mitochondria were prepared from just confluent cultures as previously described [8]. Proteins were then extracted in 7M urea, 2M thiourea, 25mM DTT, 0.4% ampholyes (3-10 range) and 2% C13E10 for a better solubilization of membrane proteins [10]. As protein concentration could not be determined in this mixture, a small aliquot of the mitochondrial preparation was extracted in the same mixture except for substitution of the 2% C13E10 detergent by 4% CHAPS. With the latter mixture, protein concentrations could be determined by a Bradford-type assay. The protein extracts were stored at -20°C until use.

### Two-dimensional electrophoresis

Two-dimensional electrophoresis was performed with immobilised pH gradients for isoelectric focusing. Home-made linear 4-8 or 3-10.5 gradients were used [11] and prepared according to published procedures [12]. IPG strips were cut with a paper cutter, and rehydrated in 7M urea, 2M thiourea, 4% CHAPS, 0.4% carrier ampholytes (3-10 range), containing either 20mM DTT (4-8 gradients) or 5mM Tris cyanoethyl phosphine (purchased from MolecularProbes, for 3-10.5 gradients) [13]. The protein sample was mixed with the rehydration solution in the case of 4-8 gradients, or cup-loaded at the anode for 3-10.5 gradients. Isoelectric focusing was carried out for a total of 60000 Vh. After focusing, the strips were equilibrated for 2 x 10 minutes in 6M urea, 2% SDS, 125 mM Tris-HCl pH 7.5 containing either 50mM DTT (first equilibration step) or 150mM iodoacetamide (second equilibration step). The equilibrated strip was loaded on the top of a 10% polyacrylamide gel, and submitted to SDS PAGE (10% gel) at 12W/ gel. Two different systems were used in the separation gel [14], allowing to zoom the optimal separation window to various Mw ranges.
After migration, the gels were stained either with silver nitrate for 2D gels with a pH 4-8 gradient [15], or with ammoniacal silver for 2D gels with a pH 3-10.5 gradient [16]. Quantitative gel analysis was performed on the silver-stained gels with the Melanie II software (Genebio, Geneva, Switzerland). The experiments were performed in triplicate, starting with different cell batches. Several gels were made for each culture, in order to select gels with very close detection signal levels for quantitative analysis. This allowed us to keep the gel analysis parameters constant for better reproducibility. In the three separation systems used, only the changes reproduced in the three pairs of samples were taken into account. Preparative gels for spot excision for MS were stained with a home-made fluorescent ruthenium complex [17].

**Mass spectrometry**

*In gel digestion :*

Excised gel slice rinsing was performed by the Massprep (Micromass, Manchester, UK) as described previously [17]. Gel pieces were completely dried with a Speed Vac before digestion. The dried gel volume was evaluated and three volumes trypsin (Promega, Madison, US) 12.5ng/$\mu$l freshly diluted in 25mM $NH_4HCO_3$, were added. The digestion was performed at 35°C overnight. Then, the gel pieces were centrifuged for 5 min in a Speed Vac and 5$\mu$l of 35% H2O/ 60% acetonitrile/ 5% HCOOH were added to extracted peptides. The mixture was sonicated for 5 min and centrifuged for 5 min. The supernatant was recovered and the procedure was repeated once.

*MALDI-TOF-MS analysis*

Mass measurements were carried out on an ULTRAFLEX™ MALDI TOF/TOF mass spectrometer (Bruker-Daltonik GmbH, Bremen, Germany). This instrument was used at a maximum accelerating potential of 20kV and was operated in reflector positive mode. Sample preparation was performed with the dried droplet method using a mixture of 0.5mL of sample with 0.5mL of matrix solution. The matrix solution was prepared from a saturated solution of a-cyano-4-hydroxycinnamic acid in H2O/ 50% acetonitrile diluted 3 times. Internal calibration was performed with tryptic peptides resulting from autodigestion of trypsin (monoisotopic masses at m/z=842.51 ; m/z=1045.564 ; m/z= 2211.105).

*MS Data analysis*

Monoisotopic peptide masses were assigned and used for databases searches using the search engine MASCOT (Matrix Science, London, UK) [18]. All proteins present in Swiss-Prot and TrEMBL were used without any species pI and Mr restrictions. The peptide mass error was limited to 50 ppm, one possible missed cleavage was accepted. Only protein identifications with a significant Mascot score (greater than 75) and the correct species (human) were selected. This strategy allowed confident identification even for low molecular weight proteins, for which the sequence coverage can be low. In some cases, the MS/MS ability of the TOF/TOF instrument was used to obtain sequence tag data. For localization predictions, the MitoProt prediction software was used [19].

## 3. Results and discussion

In our search for mitochondrial proteins differentially expressed in mitochondria devoid of mtDNA(rho0) vs. normal ones, one of the major difficulties we faced was due to contamination by non-mitochondrial proteins. Such a contamination is quite frequent [20], but variable from one mitochondrial preparation to another. Furthermore, superpurification of mitochondria, for example by free-flow electrophoresis [21] was not practicable in our case, both because of quantitative limitations but also, and more importantly, because normal and rho0 mitochondria do not have the same size and or physico-chemical parameters [22], thereby leading to different cross-contaminations. As a result, these contaminations altered in turn the quantitative measurements of proteins we could make in proteomics experiments, thereby increasing the standard deviation to ±20%. For the triplicate experiments we carried out starting from different cultures, this implied that statistically significant differences ($p<0.05$) are only obtained for quantitative differences by a factor of 1.7 or more. To increase the proteome coverage, we used both different pH gradients and different gel systems. Typical results are shown on figure 1-3, and summarized in Table 1. The various gel systems used were considered as independent. This means in turn that cross assignment of proteins from one gel system to another was not made by gel matching, but by real spot identification by mass spectrometry. In order to avoid any misinterpreation due to the comigration of several proteins in a single spot, the matching spots in gels made from control and rho0 mitochondria were always analyzed in both gels when present. This ensured that the quantitative reduction observed is a bona fide quantitative reduction, and not an apparent one due to the complete disappearance of one of the proteins composing the spot while the other comigrating proteins are kept constant.
As a matter of facts, we used our previous work on human mitochondria [20] to assign some respiratory complexes subunits that did not change significantly in rho0 mitochondria. These protein spots were re-assessed by mass spectrometry (see table 1).
On a biological point of view, several important mitochondrial functions were found altered in rho0 cells compared to normal ones

**Respiratory complexes:**
As the mitochondrial genome encodes membrane subunits of the respiratory complexes, it is obvious that important alterations in these complexes should be observed. Surprisingly enough, we did not observe a major and uniform decrease in all subunits of the respiratory complexes. For example, the subunits of complex I (NADH-ubiquinone oxidoreductase) showed very different behaviors, ranging from insignificant changes (e.g. 24kDa subunit) to major quantitative changes (above 5fold for the B16.6, 75 kDa and 23kDa subunits). This apparent discrepancy takes place while it is known that there is absolutely no functional complex I in rho0 cells [4]. However, it becomes more and more apparent that complex I is assembled via subcomplexes [23], so that some stable subcomplexes could survive in rho0 mitochondria.
A similar situation takes place in Complex IV (cytochrome oxidase), where some subunits are completely unaltered in rho0 mitochondria (e.g. CoxVa) while others are dramatically reduced (e.g. Cox Vb). Interestingly enough, the proteins involved in the assembly of complex IV, i.e. the surf proteins, were also found significantly decreased in rho0 mitochondria, thereby confirming the absence of functional complex IV.

The situation was less ambiguous for the other respiratory complexes, but in different ways. Complex II (succinate dehydrogenase); which is the only complex to be totally encoded in the nucleus, did not show any significant change in rho0 mitochondria, while complex III (ubiquinol-cytochrome c oxidoreductase) showed oppositely a significant decrease in all observed subunits, here again with some subunits (e.g. the Rieske protein) more altered than others (core proteins I and 2)
Last but not least, the ATP synthase subunits were completely unaltered in rho0 mitochondria. This is to be linked to the fact that a partially-functional ATP synthase is required for the survival of rho0 cells, not for ATP production, but as an ATPase using ATP produced by glycolysis to maintain the transmembrane pH gradient [24], [25].

**Catabolic proteins:**
The mitochondrion is also an important catabolic center. In addition to the final degradation of sugars made by the Krebs cycle, numerous catabolic steps of amino acids and lipids also take place in the mitochondrion. This catabolism often uses dehydrogenases, which normally produce reducing equivalents that are used by the respiratory chain. As this respiratory part is altered in rho0 mitochondria, it is not surprising to notice a decrease in some dehydrogenases. Here again, not all dehydrogenases were equally altered. While dihydrolipoamide dehydrogenase, hydroxyacylCoA dehydrogenase and long chain CoA dehydrogenase were significantly decreased, short chain acylCoA or isocitrate dehydrogenases were not. This means in turn that the enzymes of the Krebs cycle are not uniformly decreased in rho0 mitochondria. The best example is given by the various subunits of the oxoglutarate dehydrogenase complex, which are not altered to the same extent. It is also worth to notice that the enzyme subsequent to the oxoglutarate dehydrogenase complex in the Krebs cycle, namely succinylCoA ligase, was strongly altered in rho0 mitochondria, while the enzyme upstream, i.e. isocitrate dehydrogenase, was not.
Interestingly enough, all the subunits of the electron transfer complex, which transfers the electrons produced by various dehydrogenases to the ubiquinone, were decreased around 2fold.

**Transport systems:**
The mitochondrial inner membrane is very impermeable to many solutes, due to its barrier role in the establishment of the transmembrane pH gradient that is normally used for ATP production. This means in turn that numerous transport systems exist in mitochondrial membranes to allow selective import of useful molecules. This includes of course small ions, such as calcium for which mitochondria are a major cellular store, but also proteins, which are imported into mitochondria by a complex transport system involving both an outer membrane transport system (TOM) and an inner membrane transport system (TIM) (reviewed in [26]). It has been recently shown that some subunits of these protein import systems (e.g. TOM20, TOM34 and TIM23) are slightly decreased in rho0 mitochondria [27]. Our data complement this work, showing that TOM22 and TIM9 are decreased more than 2fold in rho0 mitochondria. We also add further information in showing that another protein import system, the metaxin system [28], [29] is altered in rho0 mitochondria.
Apart from the protein import system, we also noticed mtDNA depletion-induced changes in other mitochondrial membrane proteins. This included poorly characterized proteins such as the mitochondrial outer membrane protein 25, which function is still unknown. This also included ion transport systems, such as the calcium transporter or the stomatin-like protein 2. For the latter protein, which is not fully characterized yet, we noticed a very strong prediction for a mitochondrial localization (p= 0.99 according to the MitoProt prediction software). This protein has also been recently found in a proteomic study devoted to the inner mitochondrial membrane proteins [30], and has been assigned to ionic transport control in the membrane. As to the calcium transporter, our results correlate with recent ones showing an altered calcium signalling in Rho0 cells [31].

Within this trend of transport proteins, it is quite clear that our proteomic study is far from complete, because of the well-known flaws of 2D electrophoresis against membrane proteins [32]. However, our work already points out important changes in mitochondrial transport systems upon deletion of mtDNA.

**Mitochondrial translation system:**

The mitochondrial mRNAs transcribed from the mtDNA are translated in situ by a translation system closely related to the prokaryotic translation system. However, all the proteins which are part of this system are encoded by the nucleus and imported into the mitochondria. This includes all the mitochondrial ribosomal proteins, while the mitochondrial ribosomal RNAs are encoded by the mitochondrial DNA. Thus, the absence of mtDNA implies an absence of rRNAs in the mitochondria, and thus the impossibility of building functional ribosomes. It could be expected that mitochondrial ribosomal proteins would be absent in rho0 mitochondria. This is in fact not the case, although all mitochondrial ribosomal proteins we could visualize were indeed reduced in rho0 mitochondria. Here again, the reduction factor varies from one protein to another. The situation therefore parallels the one observed with the respiratory complexes, where the functional disappearance of a system does not mean either the disappearance of all subunits nor a coordinate reduction in all subunits.

In addition to the ribosomal proteins, other proteins working in the translation apparatus were also decreased in rho0 mitochondria. One example is the CCA-adding enzyme for the mitochondrial tRNAs, which is reduced more than 5 fold. Another example is the tryptophanyl-tRNA synthase. The fact that only one aminoacyl-tRNA synthase is detected can be interpreted either by the fact that the other aminoacyl-tRNA synthases are not significantly decreased in rho0 mitochondria, or by the fact that they are not properly detected in our system. Interestingly enough, the elongation factor EF-TU, which was found increased in other mtDNA disorders [9], was not found altered in rho0 mitochondria. Here again, this means that all components of the same system do not respond the same way. The apparent discrepancy between our results and the ones obtained on the MERRF system [9], can in fact be explained by the cellular physiology. In the MERRF system, the mitochondrial ribosomes are normal, and the alteration in one of the tRNAs produces proteins that are randomly truncated at one of the codons that are not recognized by the mutated tRNA. This means that a proportion of the proteins are either not truncated at all, or truncated in such a way that they are still at least partly functional. Thus, increasing the protein production rate (e.g. by increasing the amount of elongation factors) is a way of increasing the amount of functional protein. As the non-functional proteins are degraded by the mitochondrial proteases, the only cost incurred by the cells in MERRF is energy wasting. While this is deleterious at the organism scale because some cells cannot afford such energy losses (e.g. muscle cells and neurons), MERRF transformed cells do not grow much differently than their normal counterparts. The situation is dramatically different in rho0 cells, wher nothing can be rescued on this side. The fact that EF-TU is not altered in rho0 cells just means that this protein is normally imported but not degraded faster than usual in rho0 cells.

**Other proteins:**

Other proteins not belonging to any of the previously described functional classes were also found altered in rho0 mitochondria. This of course included some very poorly characterized proteins (see end of table 1), for which a mitochondrial localization was predicted by the MitoProt software [19]. Apart from these really putative proteins, we also detected more interesting changes. Among them, a rather poorly characterized protein, homolog even at the 3D structure level to a bacterial protein involved in the late steps of tyrosine degradation [33], and with a strong probability of mitochondrial localization.

Going on with more familiar mitochondrial protein, a slight decrease in rhodanese was observed. Rhodanese is involved in the formation of iron-sulfur (Fe-S) complexes, and the respiratory complexes are rich in Fe-S clusters. With the absence of functional respiratory complexes and thus the decrease in the amount of Fe-S clusters needed, a decrease in rhodanese is not surprising.

More surprising is the decrease in a putative hydroxysteroid dehydrogenase. However, it must be kept in mind that other steroid dehydrogenases (e.g. progesterone reductase) are found associated both with the endoplasmic reticulum and the mitochondria.

Last but certainly not least, we also found that two proteins involved in apoptosis control are decreased in rho0 mitochondria. One of them, Hax-1 is an antiapoptotic protein [34] and has been described to be associated with mitochondria [35] without having a bona fide mitochondrial import sequence. , The other one, Smac is pro-apoptotic and is a true mitochondrial protein [36]. The latter finding, i.e. the decrease of Smac in rho0 mitochondria, correlates with the lower susceptibility of 143B-rho0 cells to apoptosis [37], as this lower susceptibility has been assigned to a lower activation of caspase 3 [37], which is in fact activated by Smac [38].

**General discussion**

One of the most surprising results in our study is the fact that even for pathways which are known to be completely destroyed by the absence of mitochondrial DNA, such as respiration or mitochondrial translation, we still find variable levels of the nucleus-encoded components in Rho0 mitochondria. This can be supported by two hypotheses. First, there is no fine regulation at the nuclear transcription and cytosolic translation level, so that these proteins are made at the same time as other mitochondrial proteins that are present in pathways maintained in Rho0 mitochondria. This hypothesis is supported by the studies on the mitochondrion-nucleus dialog [39], which clearly shows that there is a raw block regulation. However, there is another hypothesis, which is that these proteins are not used only for their known function in mitochondria. In this case, their presence in Rho0 mitochondria is due to their unknown ancillary function. Furthermore, the fact that variable amounts of these proteins remain in Rho0 mitochondria can be interpretated either as a regulation due to their ancillary roles, or as an evidence of the importance of subcomplexes [23] as the major stability determinant in mitochondria. In other words, proteins engaged into stable subcomplexes survive longer and are then less prone to degradation, thereby giving higher stationary levels in mitochondria. Within this frame, the variable results obtained for mitochondrial ribosomal proteins may seem strange. However, recent work on mitochondrial ribosomes has shown that protein complexes replace in part RNA-protein complexes in the mitrochondrial ribosome [40]. Within this frame, it is likely that sub-ribosomal complexes of variable stability exist, which is consistent with the variable decreases in mitochondrial ribosomal proteins observed in Rho0 cells.

**Concluding remarks**

The large scale scope of proteomics provided a wider view of the alterations in the mitochondrial protein content caused by the absence of mitochondrial DNA. Besides obvious and well characterized trends, such as the respiratory complexes, the absence of mitochondrial DNA also alters various mitochondrial functions such as mitochondrial translation, where mitochondrially encoded components are implied (in this case RNAs). It also alters mitochondrial systems in which only nuclear-encoded proteins are present, such as various dehydrogenases and electron transfer proteins, but also import systems and apoptosis-controlling proteins. This wide range of altered proteins clearly demonstrate the pleiotropic effects of mitochondrial DNA alterations and also gives valuable markers for the study of the mitochondrial-cytosolic coordination.

**Acknowledgements**

TR wants to acknowledge personal support from the CNRS.

**Figures**

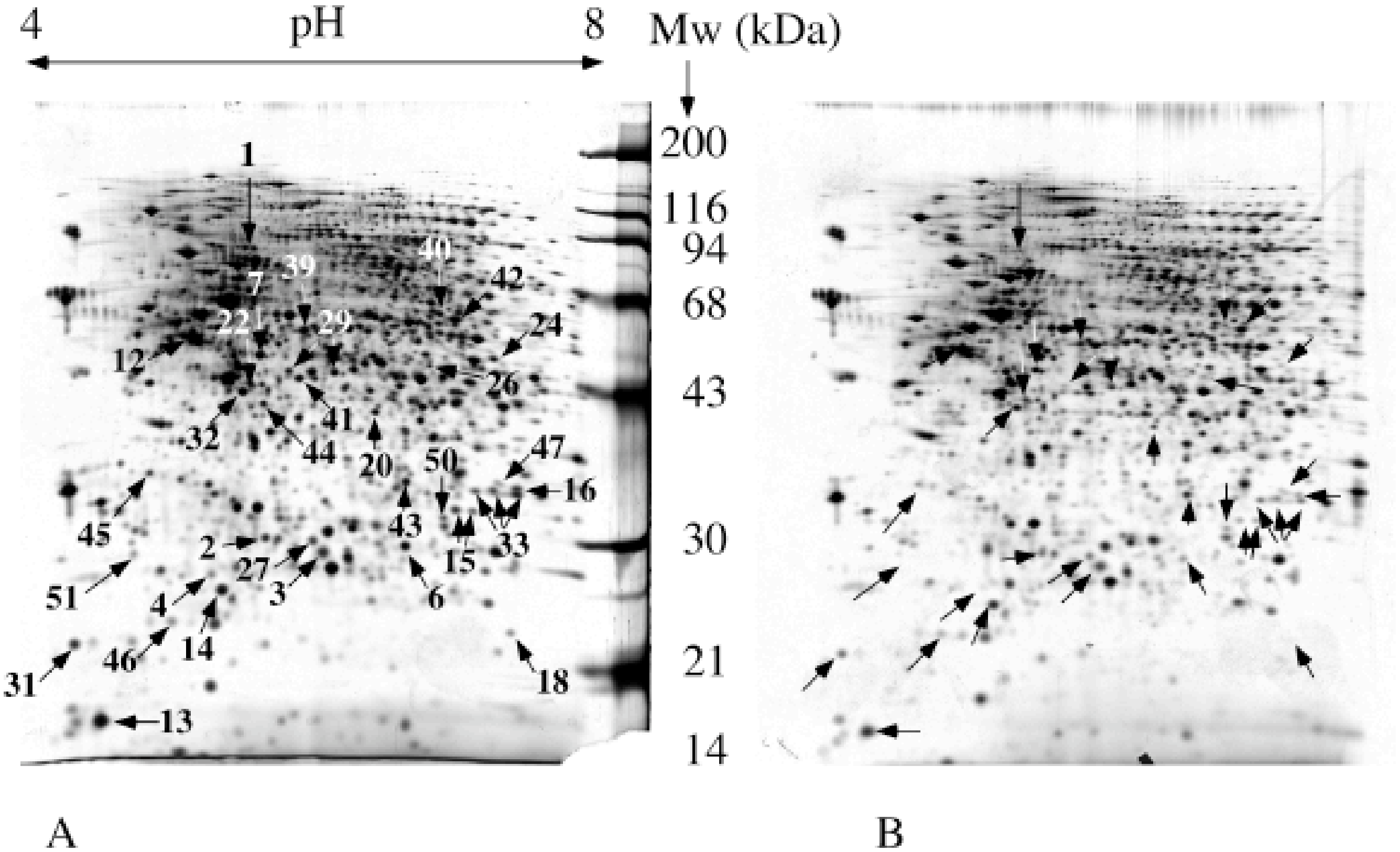

**Figure 1: Comparative analysis of normal and Rho0 mitochondria.**
150 µg of mitochondrial proteins were separated by two-dimensional gel electrophoresis. Separation parameters: pH gradient: immobilized linear pH 4 to 8. Second dimension: 10% gel at pH 8.8 (glycine system). Detection by silver staining. Only reproducible differentially expressed spots are indicated by arrows and numbered according to table 1. This means in turn that some spots appear variable in the gel pair shown, but are not consistently variable. A: normal mitochondria. B: Rho0 mitochondria

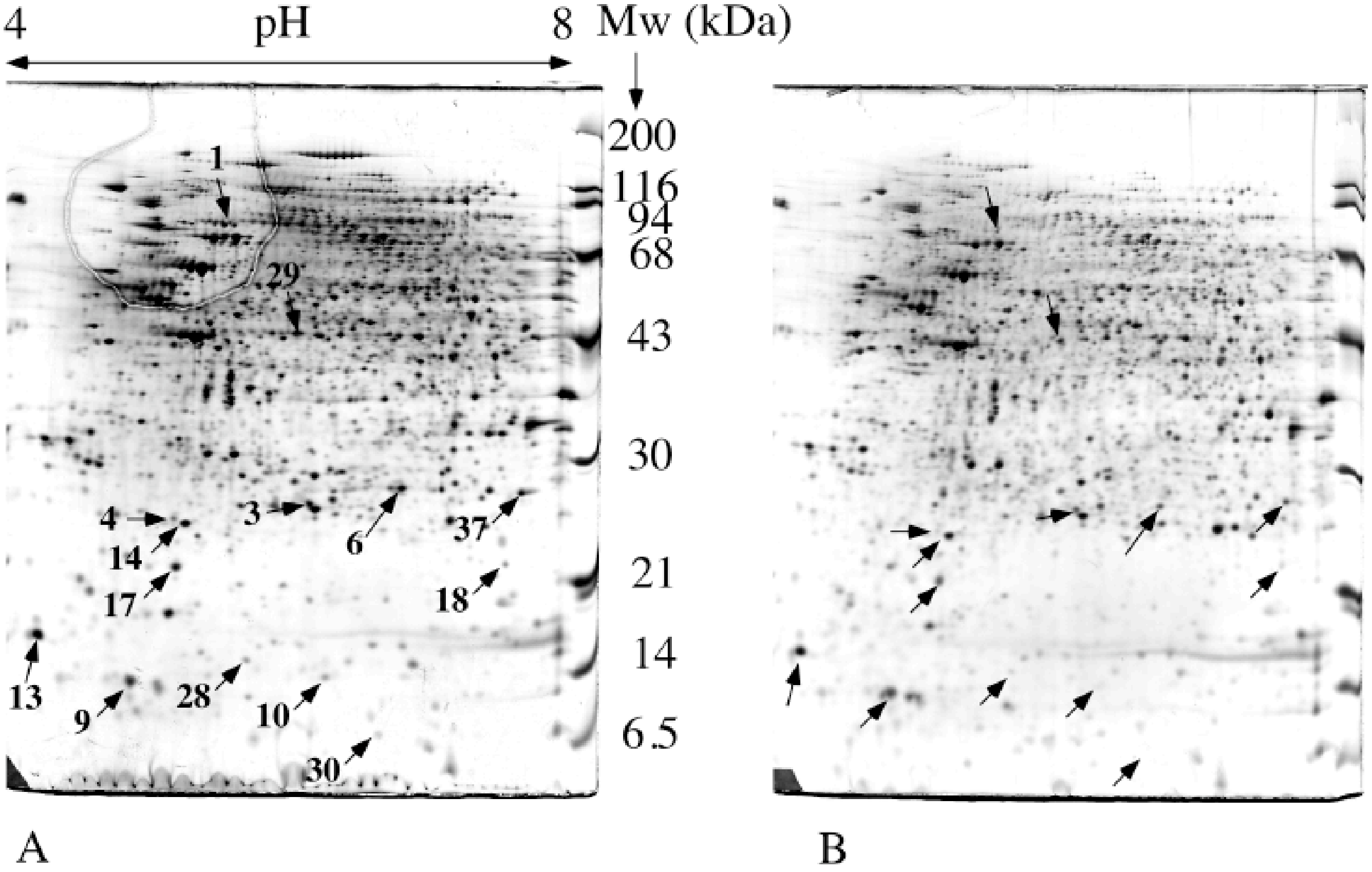

**Figure 2: Comparative analysis of normal and Rho0 mitochondria.**
120 µg of mitochondrial proteins were separated by two-dimensional gel electrophoresis. Separation parameters: pH gradient: immobilized linear pH 4 to 8. Second dimension: 10% gel at pH 8.05 (taurine system). Reproducible differentially expressed spots are indicated by arrows and numbered according to table 1. A: normal mitochondria. B: Rho0 mitochondria

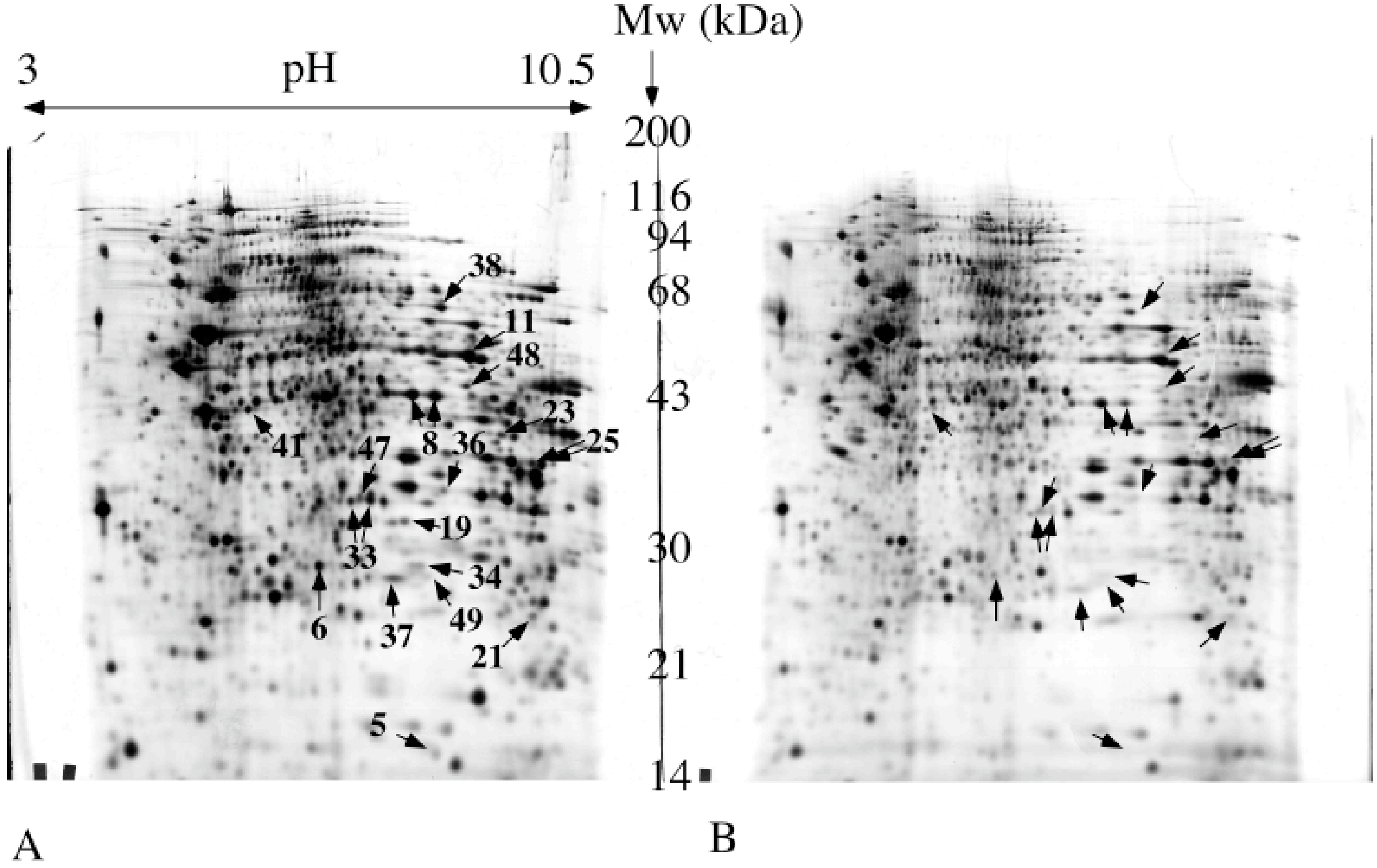

**Figure 3: Comparative analysis of normal and Rho0 mitochondria.**

150 µg of mitochondrial proteins were separated by two-dimensional gel electrophoresis. Separation parameters: pH gradient: immobilized linear pH 3 to 10.5. Second dimension: 10% gel at pH 8.8 (glycine system). Reproducible differentially expressed spots are indicated by arrows and numbered according to table 1. A: normal mitochondria. B: Rho0 mitochondria

# Table 1: summary of protein decreased in Rho0 mitochondria

| number on gels | Acc number | name | pI | Mw (kDa) | seq cov. | xfold decr in Rho0 vs 143B | |
|---|---|---|---|---|---|---|---|
| | | respiratory complexes subunits and assembly proteins | | | | | |
| 1 | P28331 | NADH-ubiquinone oxidoreductase 75 kDa subunit | 5.42 | 77 | 42 | 6 | |
| 2 | O75489 | NADH-ubiquinone oxidoreductase 30 kDa subunit | 5.48 | 26.4 | 41 | 1.6 | |
| 3 | P19404 * | NADH-ubiquinone oxidoreductase 24 kDa subunit | 5.71 | 23,75 | 30 | 1.24 | |
| 4 | O00217 | NADH-ubiquinone oxidoreductase 23 kDa subunit | 5.10 | 20.3 | 42 | 21 | |
| 5 | Q9P0J0 | NADH-ubiquinone oxidoreductase B16.6 subunit | 8.24 | 16.5 | 20 | 5.4 | |
| 6 | P47985 | Ubiquinol-cytochrome c reductase iron-sulfur subunit | 6.3 | 21.6 | 31 | 9 | |
| 7 | P31930 | Ubiquinol-cytochrome-c reductase complex core protein 1 | 5.43 | 49.1 | 46 | 1.8 | |
| 8 | P22695 | Ubiquinol-cytochrome-c reductase complex core protein 2 | 7.74 | 46.8 | 61 | 1.7 | |
| 9 | Q8TB65 * | Cytochrome c oxidase subunit Va | 4.88 | 12.5 | 18 | 0.75 | |
| 10 | P10106 * | Cytochrome c oxidase polypeptide Vb | 6.3 | 10.6 | 24 | 6 | |
| 11 | P25705 | ATP synthase alpha chain | 8.28 | 55.2 | 47 | 1.1 | |
| 12 | P06576 | ATP synthase beta chain | 5.0 | 51.7 | 20 | 1.1 | |
| 13 | P30049 * | ATP synthase delta chain | 4.53 | 15 | 15 | 0.9 | |
| 14 | O75947 * | ATP synthase D chain | 5.22 / 4.95 | 18 / 16.8 | 24 | 1.1 | |
| 15 | O75880 | SCO1 protein homolog | 8.54 | 30.3 | 20 | 3.7 | |
| | | mitochondrial translation apparatus | | | | | |
| 16 | Q9BYD6 | Mitochondrial ribosomal protein L1 | 6.46 | 31.4 | 21 | 4.5 | |
| 17 | P52815 | 39S ribosomal protein L12 | 5.37 | 16.4 | 43 | 10 | |
| 18 | Q96GC5 | Mitochondrial ribosomal protein L48 | 7.78 | 20.6 | 36 | 4 | |
| 19 | Q9Y399 | Mitochondrial 28S ribosomal protein S2 | 7.35 | 28.3 | 41 | 20 | |
| 20 | P82650 | Mitochondrial 28S ribosomal protein S22 | 6.52 | 38.1 | 52 | 3 | |
| 21 | Q9Y3D9 | Mitochondrial ribosomal protein S23 | 8.94 | 21.7 | 55 | 2.3 | |
| 22 | Q92552 | Mitochondrial 28S ribosomal protein S27 | 5.37 | 43.7 | 27 | 2.0 | |
| 23 | P51398 | Mitochondrial 28S ribosomal protein S29 | 8.75 | 43.5 | 29 | 2.9 | |
| 24 | Q9NP92 | Mitochondrial 28S ribosomal protein S30 | 8.24 | 50.3 | 30 | 10.5 | |
| 25 | Q9UGM6 | Tryptophanyl-tRNA synthetase | 8.99 /8.29 | 37.9 / 35.1 | 31 | 2.9 | |
| 26 | Q96Q11 | tRNA-nucleotidyltransferase 1 (CCA addition) | 6.29 / 6.66 | 44.9 / 48.0 | 19 | 5.5 | |
| | | mitochondrial transport systems | | | | | |
| 27 | O75431 | Metaxin 2 | 5.9 | 29.7 | 35 | 2.1 | |
| 28 | P57105 | Mitochondrial outer membrane protein 25 | 5.85 | 15.9 | 59 | 2.5 | |
| 29 | Q9P129 | Calcium-binding transporter | 6.0 | 53.3 | 31 | 3.5 | |
| 30 | Q9Y5J7 * | Mito. import inner mb. translocase subunit TIM9 A | 6.71 | 10.3 | 30 | 5.0 | |
| 31 | Q9NS69 | Mitochondrial import receptor subunit TOM22 | 4.27 | 15.4 | 77 | 2.3 | |
| 32 | Q9UJZ1 | stomatin-like protein 2 | 5.26 | 35.0 | 64 | 1.9 | 0,99 |
| | | Energy systems | | | | | |
| 33 | P13804 | Electron transfer flavoprotein alpha-subunit | 7.1 | 32.9 | 52 | 2.4 | |
| 34 | P38117 | Electron transfer flavoprotein beta-subunit | 8.25 | 27.8 | 50 | 2.3 | |
| 35 | Q16134 | Electron transfer flavoprotein-ubiquinone oxidoreductase | 6.42 | 63.7 | 37 | 2.3 | |
| 36 | Q16836 | Short chain 3-hydroxyacyl-CoA dehydrogenase | 8.38 | 32.5 | 31 | 3.5 | |
| 37 | Q99714 | 3-hydroxyacyl-CoA dehydrogenase type II | 7.65 | 26.9 | 55 | 2.3 | |
| 38 | P49748 | Acyl-CoA dehydrogenase, very-long-chain specific | 7.74 | 66.1 | 52 | 1.74 | |
| 39 | P36957 | Dihydrolipoamide succinyltransf. (2-oxoglutarate DH cplx) | 6.06 | 41.8 | 39 | 2.1 | |
| 40 | P09622 | Dihydrolipoamide dehydrogenase | 6.35 | 50.1 | 50 | 6.1 | |
| 41 | Q9P2R7 | Succinyl-CoA ligase [ADP-forming] beta-chain | 5.64 | 43.6 | 36 | 8.6 | |
| 42 | Q9HCC0 | Methylcrotonoyl-CoA carboxylase beta chain | 6.61 | 59.6 | 49 | 5.8 | |
| 43 | Q8WVX0 | putative enoyl-CoA hydratase | 6.22 | 32.1 | 49 | 1.7 | 0,98 |
| 44 | P78540 | arginase II | 5.42 | 36;1 | 44 | 1.9 | |
| | | Miscellaneous | | | | | |
| 45 | O00165 | Hax-1 | 4.76 | 31.6 | 53 | 2.8 | |
| 46 | Q9NR28 | Smac protein | 4.93 | 22.1 | 27 | 1.8 | |
| 47 | Q16762 | Rhodanese | 6.11 | 29.0 | 40 | 1.8 | |
| 48 | Q6YN16 | Hydroxysteroid dehydrogenase | 6.31 | 42.5 | 41 | 4.45 | |
| 49 | Q6P587 | fum. Acetoacetate lyase domain cont. Protein | 6.0 | 21.7 | 38 | 3 | 0,85 |
| 50 | Q9BSH4 * | UPF0082 protein PRO0477 | 6.98 | 29.9 | 17 | 2.2 | 0,98 |
| 51 | Q96EH3 | Chromosome 7 open reading frame 30 | 4.87 | 21.6 | 27 | 7 | 0,98 |

*: proteins for which MS/MS data were obtained and confirmed the identification